\documentclass[aps, prl, amsmath, amssymb, preprintnumbers, twocolumn, superscriptaddress]{revtex4}
\usepackage{amssymb, amsmath, amsthm}
\usepackage{color}
\usepackage[usenames,dvipsnames,svgnames,table]{xcolor}
\usepackage{graphicx}
\usepackage{amssymb}
\usepackage{ulem}
\usepackage{comment}
\usepackage{hyperref}
\usepackage{esint}
\usepackage{amsmath}

\newcommand{\mytitle}{Topological fractional pumping with alkaline-earth(-like) ultracold atoms}

\begin{document}

\title{\mytitle}

\author{Luca Taddia}
\affiliation{Scuola Normale Superiore, I-56126 Pisa, Italy}
\affiliation{CNR - Istituto Nazionale di Ottica, Sede Secondaria di Sesto Fiorentino, I-50019 Sesto Fiorentino, Italy}

\author{Eyal Cornfeld}
\affiliation{Raymond and Beverly Sackler School of Physics and Astronomy, Tel-Aviv University, IL-69978 Tel Aviv, Israel}

\author{Davide Rossini}
\affiliation{NEST, Scuola Normale Superiore $\&$ Istituto Nanoscienze-CNR, I-56126 Pisa, Italy}

\author{Leonardo Mazza}
\affiliation{D\'epartement de Physique, Ecole Normale Sup\'erieure / PSL Research University, CNRS, 24 rue Lhomond, F-75005 Paris, France}

\author{Eran Sela}
\affiliation{Raymond and Beverly Sackler School of Physics and Astronomy, Tel-Aviv University, IL-69978 Tel Aviv, Israel}

\author{Rosario Fazio}
\affiliation{ICTP, Strada Costiera 11, I-34151 Trieste, Italy}
\affiliation{NEST, Scuola Normale Superiore $\&$ Istituto Nanoscienze-CNR, I-56126 Pisa, Italy}

\begin{abstract}
Alkaline-earth(-like) ultracold atoms, trapped in optical lattices and in the presence of an external gauge field, can stabilise Mott 
insulating phases characterised by density and magnetic order. We show that this property can be used to realise a {\it topological fractional  pump}. 
Our analysis is based on a many-body adiabatic expansion and on time-dependent matrix-product-states numerical simulations. 
We characterise the pumping protocol by including both finite-size and non-adiabatic corrections.
For a specific form of atom-atom interaction, we present an exactly-solvable model of a fractional pump. 
Finally, the numerical simulations allow us to thoroughly study a realistic setup amenable of an experimental realisation.
\end{abstract}

\maketitle

\paragraph{Introduction.}
Since the invention of the Archimedean screw, it has been known that matter and energy can be transported, or \textit{pumped}, without 
imposing any external bias, by a  periodic modulation of some system parameters. Investigations of pumping in quantum  
systems encompass a wide range of phenomena and applications, from the definition of novel current standards~\cite{Pekola2008},
to the diagnostics of many-body quantum states~\cite{Avron2003}. In the adiabatic limit, quantum pumping becomes \textit{geometric}, meaning that it is related to the Berry phase (or its non-Abelian generalisation) accumulated during the cycle~\cite{Xiao2010}.

In his pioneering work, Thouless~\cite{Thouless1983} showed that in some one-dimensional (1D) insulating systems the pumped charge may be quantised to an integer number. The experimental demonstration of such pump had to wait for three decades till its realisation with cold atoms~\cite{Nakajima2016,Lohse2016}. 
Can the presence of a strong atom-atom interaction allow for the creation of pumps where, after one cycle, a \textit{fractional} charge is pumped? 
And, can this be exactly quantised even in 1D? 
Certain aspects of these possibilities were discussed, based on static analysis of the Hamiltonian spectral flow, in Ref.~\cite{Grusdt2014} for specific bosonic Chern insulators models, and in Ref.~\cite{Zeng2015} for fermionic ladders, where quantisation is not expected in true 1D systems due to gapless edge modes. 
A general discussion of interacting 1D models, where robust quantisation emerges, remains absent.

\begin{figure}[t]
  \includegraphics[width=0.56\columnwidth]{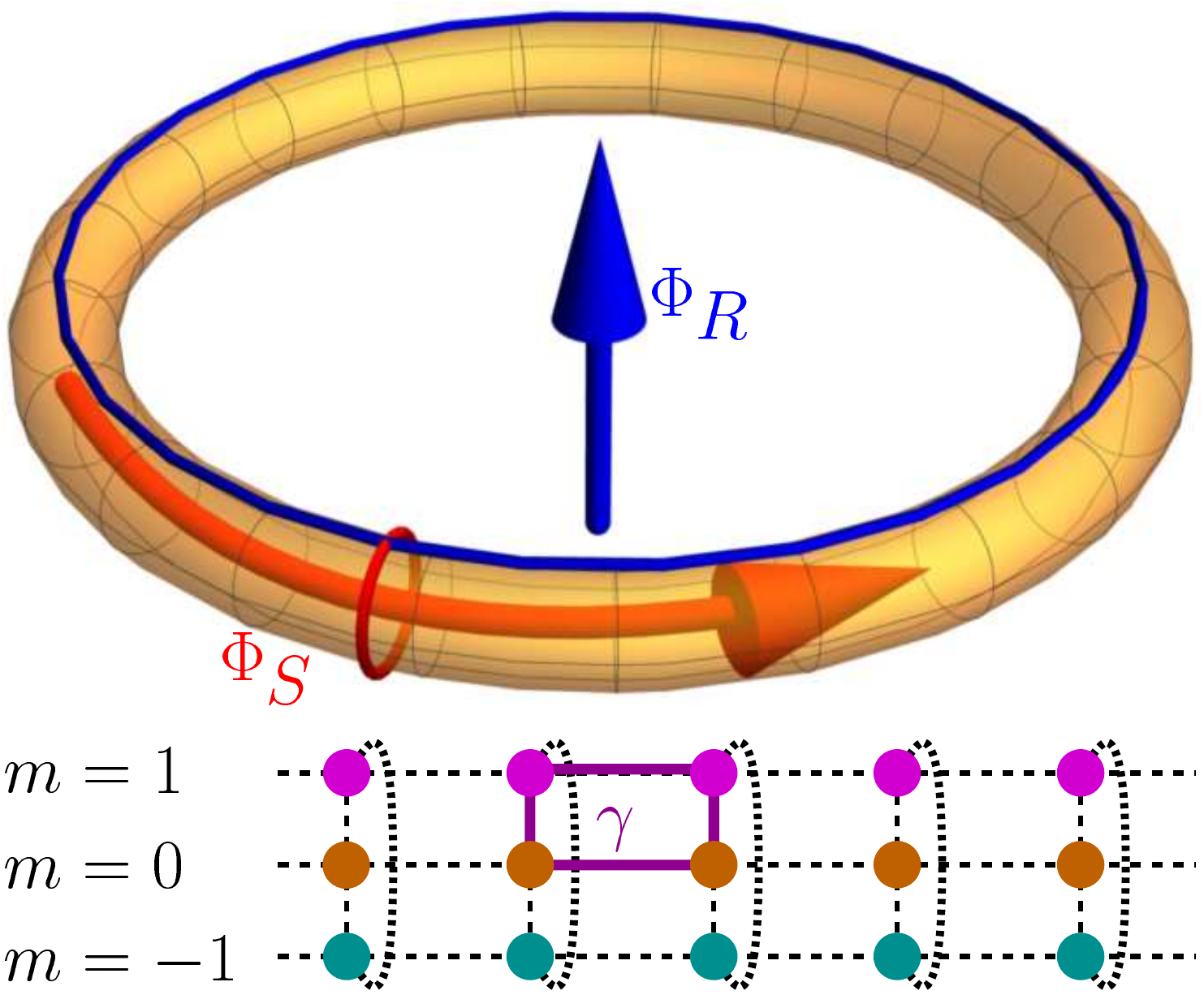}
  \includegraphics[width=0.42\columnwidth]{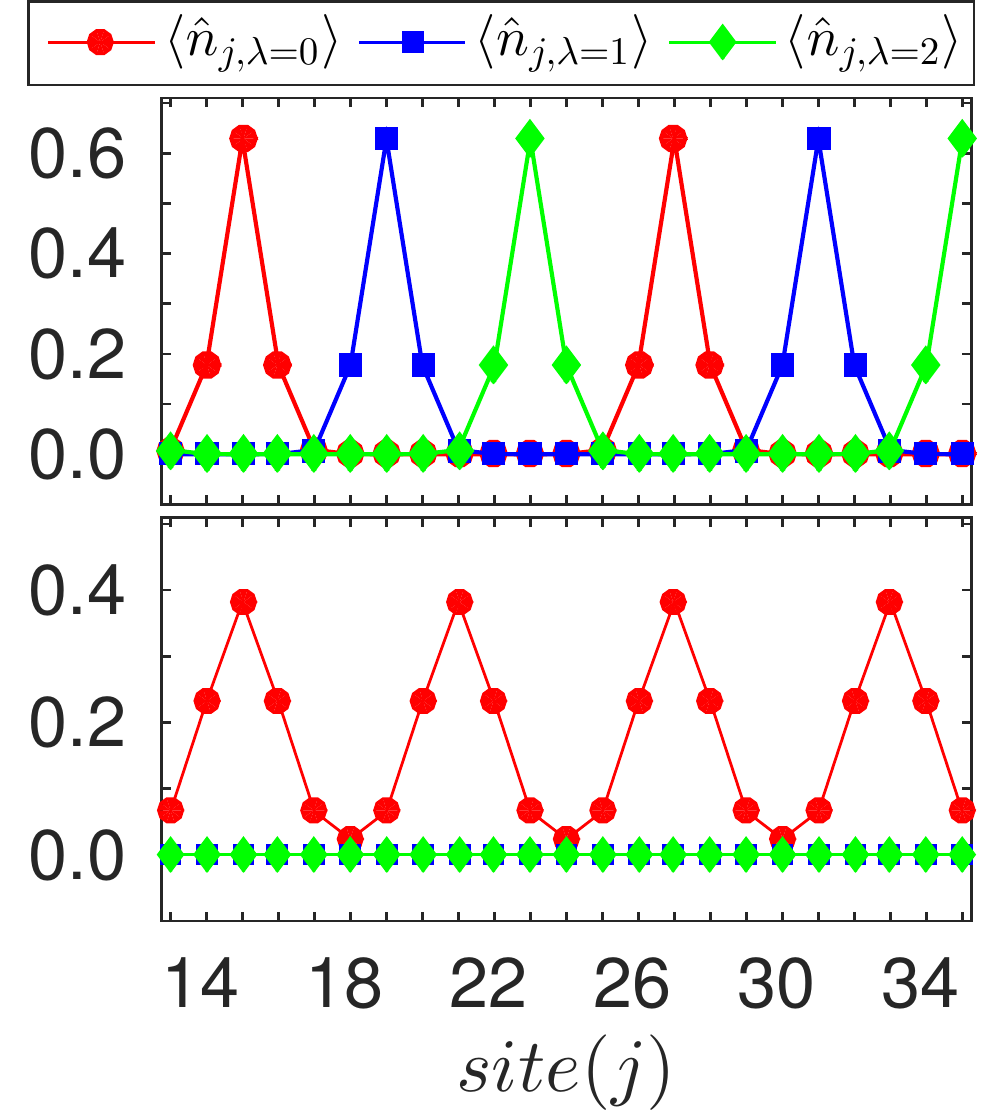}
  \caption{(Color online)
    	Left: The geometry of the system with periodic boundary conditions is a thin torus with a long real dimension and a short synthetic one.
    	Two fluxes $\Phi_R$ and $\Phi_S$ pierce the two non-contractible circles. The system is realised by coupling three atomic spin states with Raman couplings. A flux $\gamma$ is enclosed in every synthetic plaquette.
    	Right: At fractional fillings $\nu = 1/2$ (top) and $\nu = 1/3$ (bottom), Hamiltonian~\eqref{Ham} displays magnetic crystals. 
	The nature of these states is best analysed taking the Fourier transformation $\hat d_{j,\lambda} = (2 \mathcal I+1)^{-1/2} \sum_m 	e^{2 \pi i m \lambda / (2 \mathcal I +1)} \hat c_{j,m}$ (accordingly, $\hat n_{j,\lambda} = \hat d_{j,\lambda}^\dagger \hat d_{j,\lambda}$).
 	The simulation parameters coincide with those of Fig.~\ref{PumpCyl}.
  }
  \label{disegni}
\end{figure}

The recent progresses in the manipulation and in the understanding of alkaline-earth(-like) gases~\cite{Fukuhara2007, Gorshkov2010, Cazalilla2014, Zhang2014, Pagano2014, Mancini2015, Barbarino2015, Cooper2015, Hofrichter2016, Barbarino2016, Nakajima2016}
paves the way for assessing their role of strong candidates to measure fractional pumping. The experimental observations of  
interaction effects~\cite{Pagano2014} and of persistent currents induced by an external gauge field~\cite{Mancini2015} motivate a 
detailed dynamical analysis of the experimental system in order to understand under which conditions a fractional pump can be realised. In this Letter we accomplish this task. 

By means of time-dependent algorithms based on matrix-product states and of a many-body adiabatic expansion, we show that it is possible to realise a fractional topological pump with alkaline-earth gases. We characterise the dynamical protocol by addressing both finite-size and non-adiabatic corrections, as well as the role of the trap. 
We finally discuss the detection of fractional pumping through centre-of-mass dynamics~\cite{Wang2013,Nakajima2016}.  
By elucidating the relation between the pump and the many-body states,
we clarify its connection to the fractional quantum Hall states~\cite{Bergholtz2008, Prange1990}.

\paragraph{Model.} We consider a one-dimensional gas of alkaline-earth(-like) fermions with $2 \mathcal I+1$ internal states~\cite{Gorshkov2010, 
Pagano2014}, coupled through Raman beams~\cite{Celi2014, Mancini2015, Barbarino2015, Cooper2015} (the coupling between $m = \pm \mathcal I$ 
requires a multi-photon transition). The (time-dependent) Hamiltonian reads:
\begin{eqnarray}\label{Ham}
	\hat H(\tau)  \hspace{-0.12cm} &=& \hspace{-0.08cm}\sum_{j,m} \left[t(\tau) \hat c_{j,m}^\dagger\hat c_{j+1,m}
	+\Omega_{j,m}(\tau)\hat c_{j,m}^\dagger\hat c_{j,m+1}+\mbox{H.c.}\right] \nonumber \\
	&+&  \hspace{-0.2cm}  \sum_{i,j,m,m'} U_{i,j}^{m,m'}\hat n_{i,m}\hat n_{j,m'}
	+ w_0 \sum_{j,m}  \left(j-j_0\right)^2 \hat n_{j,m}  ; \quad
\end{eqnarray}
$\hat c^{(\dagger)}_{j,m}$ annihilates (creates) a fermion of spin $m = -\mathcal I, \ldots, \mathcal I $ at site $j=1,\ldots,L$ ($\hat n_{j,m}=\hat c_{j,m}^\dagger\hat c_{j,m}$ and $\hat n_j = \sum_m \hat n_{j,m}$).
The first terms represent the atomic hopping and the spin flip, with coupling $t(\tau) = -t \exp \big[ i \Phi_R(\tau) /L \big]$ and $\Omega_{j,m}(\tau) 
= \Omega \exp \big[-i \gamma j +i \Phi_S(\tau)/(2\mathcal{I}+1) \big]$ respectively. The hopping amplitude is $t$, whereas $\Omega$ is the 
amplitude of the Raman coupling~\cite{footnote0}.
The time-dependent phases $\Phi_{R/S}(\tau)$ ($\tau$ is the time)  lead to pumping~\cite{Cooper2015}; $\gamma$ induces a static gauge potential.
The atom-atom interaction in the second term is typically short-ranged and SU$(2 \mathcal I+1)$ invariant~\cite{Cazalilla2014, Capponi2016}. We consider the form $U_{i,j}^{m,m'} = \frac{U}{2}[1-\delta_{m,m'}] \delta_{i,j}+ V\delta_{j,i+1}$ with on-site and nearest-neighbour couplings $U$ and $V$, respectively. 
The last term in Eq.~\eqref{Ham} represents the harmonic confinement, of strength $w_0$, centred around $j_0=(L+1)/2$.

The Hamiltonian in Eq.~\eqref{Ham} can be interpreted as a model of spinless fermions on a $L\times(2\mathcal{I}+1)$ cylinder (torus) for 
open (periodic) boundary conditions~\cite{Boada2012} pierced by a flux per plaquette $\gamma$~\cite{Celi2014}. 
This mapping is known as \textit{synthetic gauge field in synthetic direction.}
When the density $n$ and $\gamma$ are commensurate~\cite{Oreg2014}: 
\begin{equation}
\nu= \frac{2 \pi n}{(2\mathcal{I}+1) \gamma}= \frac pq,
\end{equation}
and interactions are strong and long-range enough, the ground state is a gapped magnetic crystal~\cite{Barbarino2015}, see Fig.~\ref{disegni}  right. 
Charge and spin order appear; each magnetic crystal is $q$-fold degenerate, because of the existence of different equivalent configurations of the density-wave.
The state is related to the quantum Hall effect in the thin-torus limit, where the system size in one direction is not much longer than the magnetic length~\cite{Prange1990, Bergholtz2008}.
Indeed, in the synthetic-dimension framework, $\nu$ is the ratio between the number of particles and the number of fluxes piercing the surface of the system; moreover, the synthetic length, $2\mathcal I +1$, is comparable to the magnetic length.

\paragraph{Topological fractional pump.}
The phases $\Phi_{R,S}(\tau)$ in Eq.~\eqref{Ham} represent time-dependent fluxes (we first consider periodic boundary conditions and no trap $w_0=0$)
piercing, respectively, the real and the synthetic circles (see Fig.~\ref{disegni}, left). 
By adiabatically varying them from 0 to $2\pi$ 
in time $T$, the magnetic crystals realise a {\it fractional Thouless pump}. This scheme follows Laughlin's argument in the quantum Hall effect~\cite{Laughlin1981}.

In order to characterise the pump, we perform a next-to-adiabatic expansion of the Schr\"{o}dinger equation $T^{-1}i\partial_s |\Psi(s)\rangle=\hat{H}(s)|\Psi(s)\rangle$ with scaled time $s=\tau/T$~\cite{Ortiz2008-2014}. The magnetic crystals at filling $\nu = p/q$ are $q$-fold degenerate on a torus: for any instantaneous 
energy level $E_n(s)$, we must consider the multiplet $|n^h(s)\rangle$ satisfying $\hat{H}(s)|n^h(s)\rangle=E_n(s)|n^h(s)\rangle$;
$h = 1, \ldots, q$ labels the degenerate eigenstates, $n=0$ for the ground states. The time dependent wavefunction is $|\Psi^h(s)\rangle \simeq |\Psi^h(s)\rangle_0 + 
|\Psi^h(s)\rangle_1 +...$, with $\|| \Psi^h(s)\rangle_m \| = \mathcal{O}(T^{-m})$, and $|\Psi^h(s)\rangle_0$ the instantaneous wavefunctions.

We focus on pumping along the real direction, with $\Phi_S(\tau) =2\pi \tau/T$ and $\Phi_R$ kept constant in time.
The real-space current is given by $ \hat{J}_R(s)=\partial_{\Phi_R} \hat{H}(s)$ and the charge (or, more correctly, \textit{density}) pumped in one period is $Q_R^{(h)}=T \int_0^1 ds \langle \hat{J}_R(s)\rangle $.
It can be expanded in powers of $T^{-1}$ using the expansion for $|\Psi^h(s)\rangle$
(see also~\cite{Mottonen2006, Brosco2008}).
The zero-th order term describes a persistent current in the ground state;
by the symmetries of the Hamiltonian, it vanishes when $\Phi_R=0$.
Thus, $Q_R$ descends from the first-order correction of the adiabatic expansion.
Higher-order terms generate the non-adiabatic corrections $\delta Q_R$.
Details with their lengthy explicit expressions are presented
in the Supplementary Materials~\cite{SuppMat}.

It is convenient to average over the $q$ states (pumping is independent of the initial eigenstate), 
$Q_R=q^{-1}\sum_hQ_R^{(h)}$, obtaining~\cite{SuppMat}:
\begin{equation}\label{QR}
  Q_R = i\int_0^{2\pi} d\Phi_S\frac{1}{q}\sum_h \left[\langle \partial_{\Phi_S} 0^h|\partial_{\Phi_R} 0^h\rangle-\langle
    \partial_{\Phi_R} 0^h|\partial_{\Phi_S} 0^h\rangle \right].
\end{equation}
Due to the  degeneracy of  the ground space, the pumped charge $Q_R$ is related to the Wilczek-Zee (WZ)~\cite{Wilczek1984} 
curvature matrix
\begin{align}
  [\Omega_{WZ}]^{h'h''}=&i\big[\langle \partial_{\Phi_S} 0^{h'}|\partial_{\Phi_R} 0^{h''}\rangle +\nonumber \\
    &+\sum_{h}\langle\partial_{\Phi_S} 0^{h'}| 0^{h}\rangle \langle 0^{h}|\partial_{\Phi_R} 0^{h''}\rangle \big]+\mathrm{H.c.}
\end{align}
by \textit{averaging} over the constant flux $\Phi_R$:
\begin{equation}\label{QWZ}
  \bar{Q} = \frac{1}{2\pi}\int_0^{2\pi} Q_R d\Phi_R
  = \frac{1}{q}\oiint\frac{d^2\Phi}{2\pi}\sum_h[\Omega_{WZ}]^{hh} = \frac{\mathcal C_1}{q}.
\end{equation}
The closed surface integral over the trace of the curvature, $\mathcal C_1 $, is the integer topological invariant known as the non-Abelian first Chern number~\cite{Varadarajan}. 

Following Ref.~\cite{ThoulessNiu1985}, we can relate it
to the pumped charge $Q_R$, which is not averaged over $\Phi_R$:
\begin{equation}
  Q_R=\bar{Q}+\mathcal{O}(L^{-1}).
\end{equation}
$Q_R$ approaches the quantised value only in the thermodynamic limit
but averaging over $\Phi_R$ cancels the error even at finite size.
A similar analysis holds for pumping $Q_S$ along the synthetic direction
with $\Phi_R(\tau)=2\pi\tau/T$ and $\Phi_S$ kept constant.
Due to the intrinsically short length of the synthetic direction, finite-size corrections of order $\mathcal{O}((2 \mathcal{I}+1)^{-1}) = \mathcal{O}(1)$ are sizeable~\cite{Zeng2015}.
Again, averaging over $\Phi_S$ yields the exact quantised value~\cite{SuppMat}.

Next-order terms in the adiabatic expansion allow to evaluate the non-adiabatic corrections, which are important at finite $T$~\cite{SuppMat}:
\begin{equation}\label{Tvar:text}
  \delta Q_R = \sum_{n>0}\frac{A_n+B_n\cos(\bar\Delta_{n0}T+\varphi_n)}{\bar{\Delta}_{n0}T};
\end{equation}
where $\bar{\Delta}_{n0}=\int_0^1 [E_n(s)-E_0(s)] ds$, and $A_n$, $B_n$, $\varphi_n$ are computable constants.

\paragraph{Exactly-solvable model of a fractional pump.}
For a specific form of the interaction, we can study the pumping exactly and prove that $\mathcal C_1 / q$ is a fractional number.
We move to open boundary conditions and consider a hard-core interaction of range $\xi$ in the real dimension, namely $U_{i,j}^{m,m'} \to \infty$ for $|i-j| \leq \xi$ and zero 
otherwise. In this case, we can eliminate the $\xi$ empty rungs to the right of the first $N-1$ particles along the real dimension, 
and write an effective model defined on a system of reduced length $L'=L-(N-1) \xi$~\cite{Sela2011, Cornfeld2015, SuppMat}. 
Hard-core particles hop in the reduced lattice similarly to before: this part of the Hamiltonian is formally unchanged~\cite{SuppMat}. The spin-flip part is also formally unchanged, 
but the site-dependent phase $\gamma j$ is replaced by $\gamma j +\gamma \xi \sum_{j'=1}^{j-1} \hat n_{j'}$, thus introducing a 
non-local Hamiltonian term. When the flux satisfies $\gamma \xi = 2 \pi a$, with $a \in \mathbb N$, the model turns local again; from now on, we consider this case. Through this flux-attachment transformation~\cite{fluxattach}, we obtain a model
characterised by hard-core on-site interactions with modified filling $\nu' = \nu L/L' = 1/(\nu^{-1} - N_\Phi)$, where $N_\Phi = \frac{\gamma}{2 \pi} (2 \mathcal I +1) \xi = (2 \mathcal I+1)a$ is the number of flux quanta attached to each hard-core particle.

This transformation allows to map the fractional $\nu=p/(1+N_\Phi p)$ state to the integer $\nu'=p$. In this case, we can exactly relate the topological integral $\bar{Q}$ in Eq.~\eqref{QWZ} to $\bar Q'$, the topological integral of the effective model~\cite{SuppMat}: 
$\bar{Q}= \bar Q' / q$.
In order to compute $\bar Q'$, we observe that for zero on-site interaction the system is in a gapped phase characterised by $\bar Q' = p$~\cite{footnote}.
Since we expect that repulsive interactions stabilise the gap of the system, no phase transition is envisioned for $U \to \infty$ and thus $\bar Q' = p$ also for our effective model~\cite{SuppMat}. 
Concluding, $\bar Q = p/q$.

It is intriguing to observe that the pumped charge is related to a many-body Chern number, the topological invariant 
usually employed to characterise the wavefunctions of the quantum Hall effect~\cite{ThoulessNiu1985, Tao1986}.
Since the magnetic crystals are related to the thin-torus limit of the two-dimensional quantum Hall effect~\cite{Barbarino2015}, this shows that 
in this limit the topological invariant is not lost. However {\it these states do not display any topological order}, not even with symmetry protection. 
This is reflected by the existence of crystals at all fillings $\nu = p/q$, whereas fractional Hall states only appear for odd $q$.  
Thus, the fractional pumping property does not require the existence of a related gapped state in the fractional quantum Hall effect.

\begin{figure}[t]
  \includegraphics[width=\columnwidth]{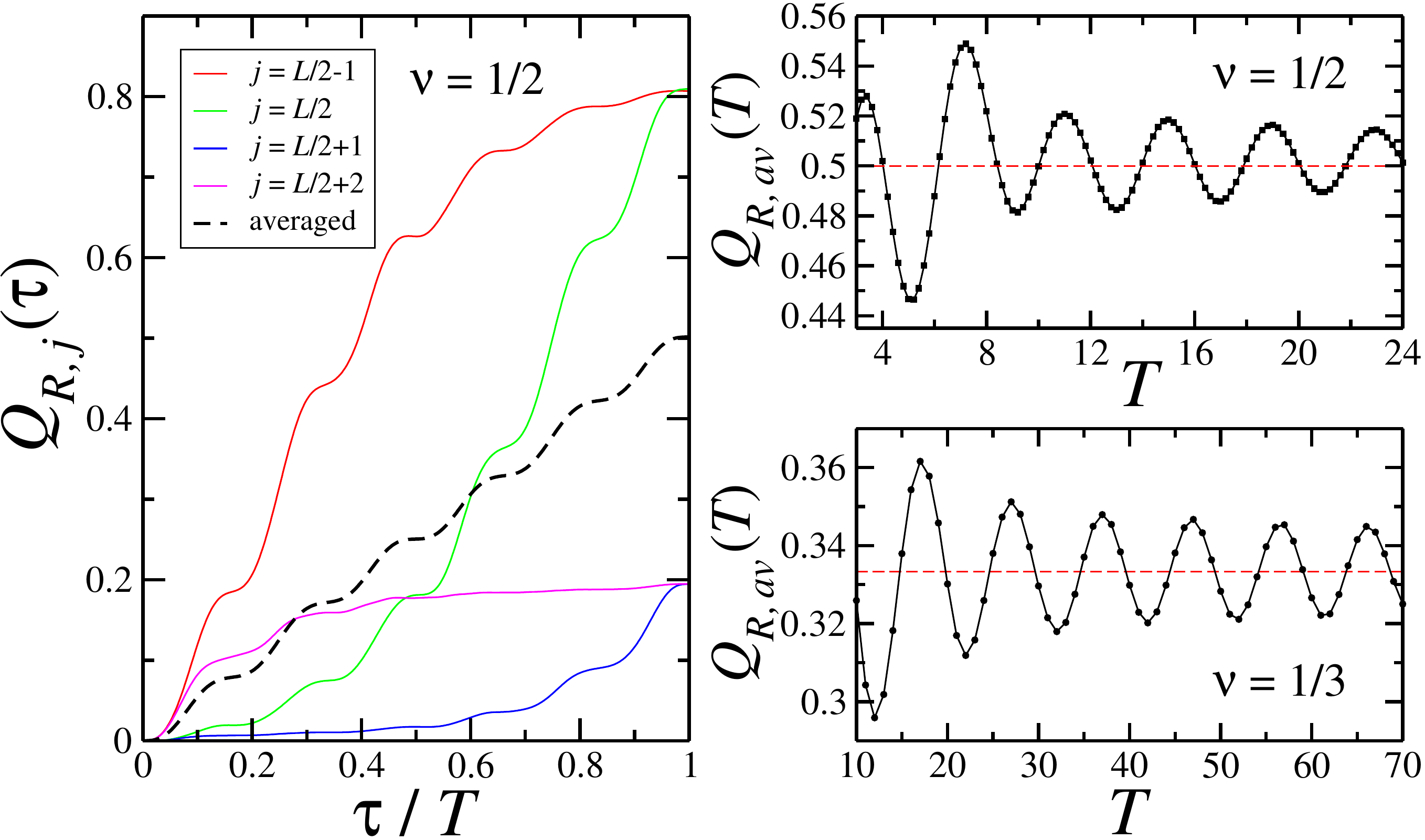}
  \caption{Pumped charge after one period for a system with hard-wall boundary conditions, $L = 48$,
    $\gamma=\pi/3$, $U/t=\infty$, $V/t=10$; $\Omega/t=1$ for $\nu=1/2$ and $\Omega/t =0.25$ for $\nu=1/3$.
    Left panel: space-resolved pumped charge $Q_{R,j}(\tau)$ for $\nu = 1/2$ and $T=24$.
    Right panels: pumped charge $Q_{R,av}(T)$ as a function of the pump period $T$ for $\nu=1/2$ (top)
    and $\nu =1/3$ (bottom). Dashed lines show the quantised values.}
  \label{PumpCyl}
\end{figure}

\paragraph{Numerical simulations.}
In order to make explicit predictions, we perform numerical simulations for $\mathcal I = 1$ and $\gamma=\pi/3$, focusing on the fractions $\nu=1/2$ (with no counterpart in the 2D Hall physics), and
$\nu=1/3$ (see Fig.~\ref{disegni}, right). In both cases, a nearest-neighbour interaction is necessary in order to stabilise the crystal~\cite{Barbarino2015}. The simulations are based on the time-dependent matrix-product-states approach: the time-evolution operator is decomposed using a fourth-order Trotter approximation with time steps between $10^{-3}t^{-1}$ and $10^{-2}t^{-1}$, keeping a bond link up to $200$~\cite{Schollwock2011}.
Open boundary conditions and pumping in the real dimension ($\Phi_R=0$) are considered; we first ignore the trap.

In the left panel of Fig.~\ref{PumpCyl}, for fixed $T$, we show the space-resolved pumped charge $Q_{R,j}(\tau) = \int_0^\tau {\rm d}\tau'\left\langle\psi(\tau')\right|\hat J^{(R)}_j\left|\psi(\tau')\right\rangle$, where
$\left|\psi(\tau)\right\rangle$ is
the time-evolved state and $\hat J^{(R)}_j=-it\sum_m\hat c_{j,m}^\dagger\hat c_{j+1,m}+\mbox{H.c.}$ is the current operator.
Since the ground state is inhomogeneous, it displays a dependence on $j$, with periodicity $\ell = 4$ for $\nu=1/2$ and $\ell = 6$ for $\nu=1/3$
($\ell$ is the space periodicity of the density profile; here, $\ell = n^{-1}$). In the right panel, we  show that quantisation
is restored once a spatial average over $\ell$ sites is performed. Note that the charge transferred in a fraction of the
pumping period is also quantised~\cite{Marra2015}.
In Fig.~\ref{PumpCyl}, right, we show the $T$-dependence of this averaged value $Q_{R,av}(\tau=T)$.
As predicted by Eq.~\eqref{Tvar:text}, it oscillates around the quantised values,
the amplitude of such oscillation vanishing in the adiabatic limit.

\begin{figure}[t]
  \centering
  \includegraphics[width=\columnwidth]{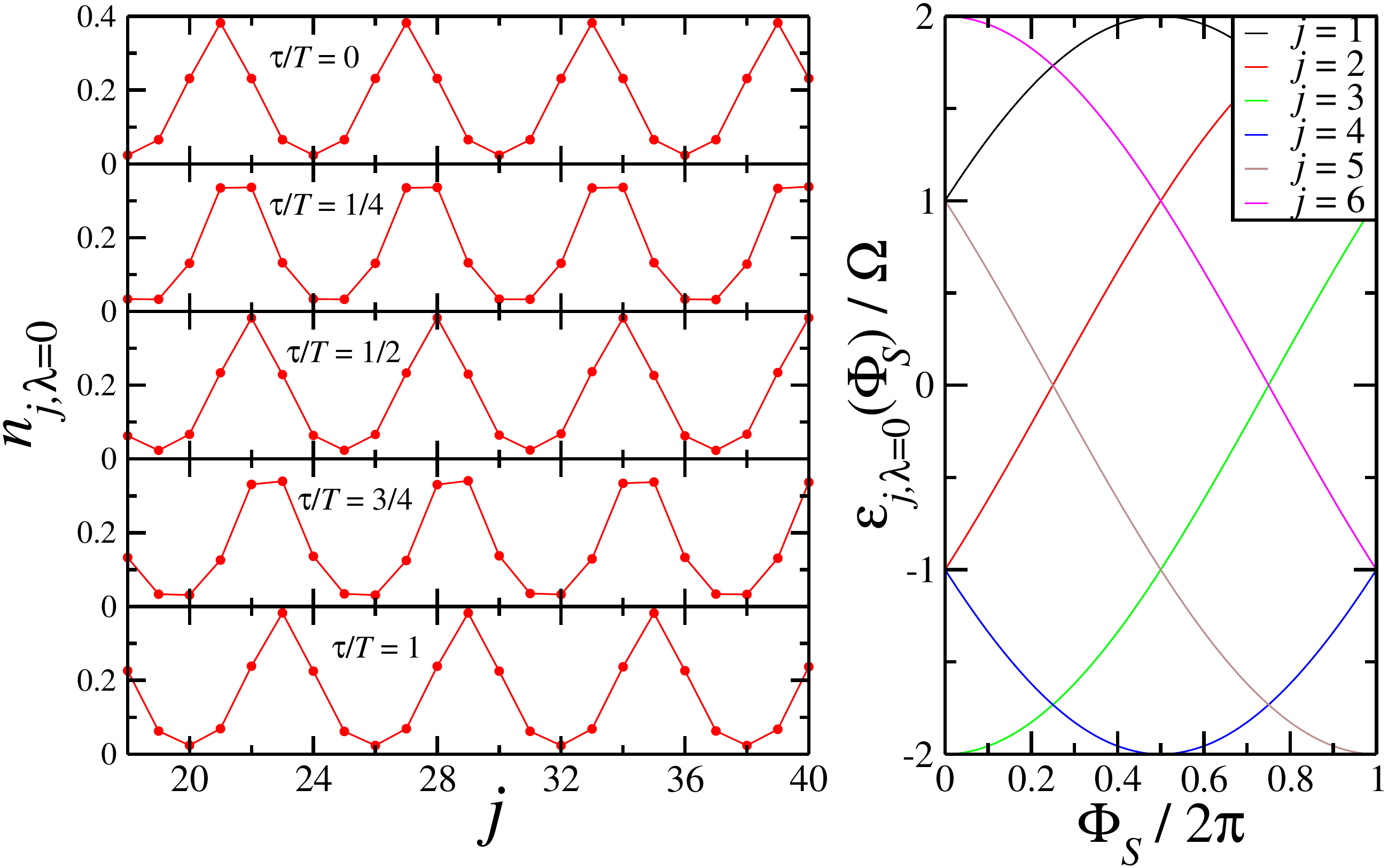}
  \caption{Left panel: time evolution of the density profile at $\nu=1/3$.
    The values of the parameters are specified in the caption of Fig.~\ref{PumpCyl}, and $T=100$.
    Right panel: flux-dependence of $\epsilon_{j,\lambda=0}\left(\Phi_S\right)$.}
  \label{TranspCyl}
\end{figure}

The quantisation of $Q_{R,av}$ can also be understood by looking at the dynamics of the magnetic crystals. In the left panel of 
Fig.~\ref{TranspCyl}, we show the time evolution of the density profiles at $\nu=1/3$: for large $T$, the particles have moved by two sites~\cite{Grusdt2014,Cooper2015}. 
In the right panel of Fig.~\ref{TranspCyl}, we plot the adiabatic time evolution of the eigenvalues of the spin-flip term in Eq.~\eqref{Ham}, $\epsilon_{j,\lambda}\left(\Phi_S\right)=2\Omega\cos\left[\frac{2\pi\lambda-
\Phi_S}{2\mathcal{I}+1} +\gamma j\right]$, for $j =1,\ldots,6$ and $\lambda=0$.  
Due to the hopping term, the crossings in the right panel of Fig.~\ref{TranspCyl} turn into avoided crossings which can be adiabatically followed for large-enough $T$.
A particle initially sitting at $j=3$ is transported to $j=4$ at $\tau=T/2$ (or $\Phi_S = \pi$) and then to $j=5$ at $\tau=T$ (or $\Phi_S = 2 \pi$). 
From the definition of $\epsilon_{j,\lambda}$ we infer that,
in a pump period, particles are adiabatically transported along
$\Delta j =2 \pi / [(2 \mathcal I+1) \gamma]$ sites. The net pumped density (the average inter-particle distance is $n^{-1}$) is given by the product $n \times \Delta j = \nu$, which is fractional, and coincides with our numerical results.

\begin{figure}[t]
\centering
\includegraphics[width=\columnwidth]{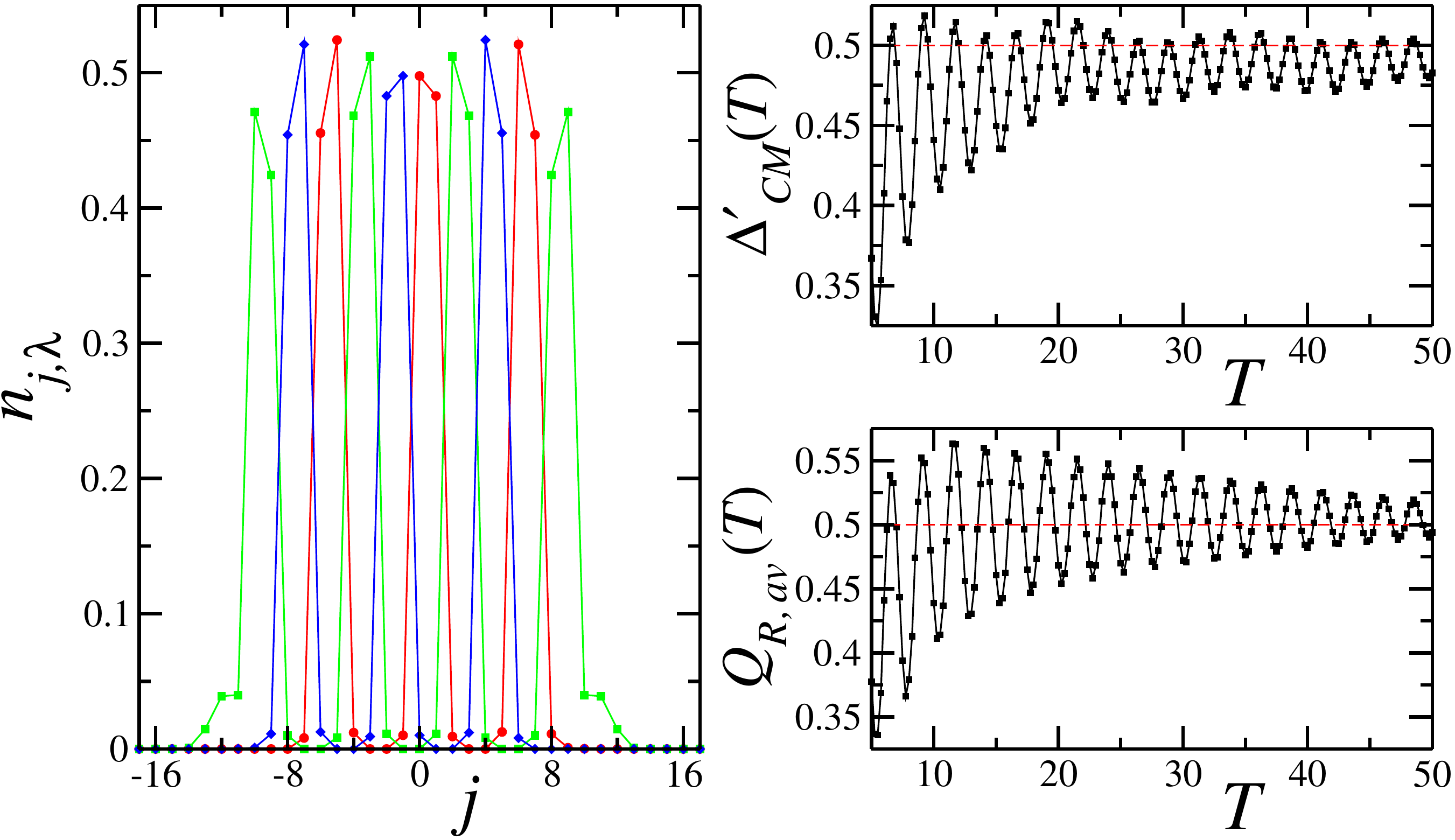}
\caption{Left: magnetic crystal with $\nu=1/2$ in a harmonic potential.
  System parameters: $\gamma=2\pi/3$, $U/t=\infty$, $\Omega/t=1$,
  $w_0/t=0.01$, $L=48$, $N=10$.
  Right: $\Delta'_{CM}(T)$ (upper panel) and $Q_{R,av}(T)$ (lower panel)
  as a function of the pump period $T$ for the same state.
  Dashed lines show the quantised values.}
\label{TrapFig}
\end{figure}

\paragraph{Experimental detection.}
In order to ascertain the possibility to measure the fractional pump, it is necessary to consider the role of a trapping potential 
and analyse measurable quantities that carry information about pumping. We switch to $\gamma = 2 \pi /3$, which allows the stabilisation 
of magnetic crystals at $\nu = 1/2$ in the experimentally relevant case of no nearest-neighbour interaction, $V=0$.
As shown in the left panel of Fig.~\ref{TrapFig}, magnetic crystals can appear in the centre of the system~\cite{Barbarino2015}.

An experimentally-viable way of estimating the pumped charge exploits the displacement of the centre-of-mass of the atomic cloud 
$\Delta_{CM}(T)$~\cite{Wang2013}. 
At low fillings, like in Fig.~\ref{TrapFig}, the system is almost entirely in a magnetic crystal and the motion of the whole cloud  can be directly related to pumping. 
In the top right panel of Fig.~\ref{TrapFig} we plot $\Delta'_{CM}(T) =\Delta_{CM}(T)/\ell $~\cite{Wang2013}.
We observe a good quantisation of $\Delta_{CM}'$, and an excellent comparison with the pumped charge $Q_{R,av}(T)$ (see Fig.~\ref{TrapFig}, 
bottom right panel), apart from the expected non-adiabatic corrections. 
The harmonic trap facilitates the pumping because of the compressible boundaries which act as source and drain leads.
The effect would be hindered by hard-wall boundaries.

Ultracold $^{173}$Yb gases with $\mathcal I=1$ have already been realised~\cite{Pagano2014, Mancini2015}, and realistic schemes
exist for the multi-photon coupling between the $m=\pm 1$ states~\cite{Celi2014, Cooper2015}.
$\Phi_S$ is a global phase associated to the spin-flip Raman beams, and can be manipulated in time~\cite{Cooper2015}. In the experiment, $\Phi_R$ can be neglected. 
Measurements of centre-of-mass displacements can be performed~\cite{Nakajima2016}, and executing several pump cycles can make the effect more detectable. 
Real-space currents can be measured by asymmetries in the spin-resolved momentum distribution function~\cite{Mancini2015, Barbarino2016},
which is measured with time-of-flight imaging.
Finally, the largest value $T = 50 t^{-1}$ in Fig.~\ref{TrapFig} corresponds to $500$ ms,
taking $t \sim 100$ Hz, which is realistic.

Concluding, we comment on two additional sources of imperfect fractional 
quantisation: (i) the existence of small metallic wings at  the edges of the crystal and (ii) the presence of multiple 
copies of the system. In a local density approximation,  this implies that part of the gas is not in a magnetic 
crystal, and thus it is not exactly pumped. On one side, this requires a careful tuning of the experimental 
parameters to lower such effects; on the other hand, our  analysis shows that imperfections induced by harmonic confinement can be negligible.

\paragraph{Conclusions.}
We have discussed the realisation of a topological fractional pump with alkaline-earth(-like) fermionic atoms. 
We have presented a detailed analysis of fractional charge quantisation, as well as of finite-size and non-adiabatic effects.
Together with an analytical adiabatic expansion, we performed extensive numerical simulations based on time-dependent matrix-product-state algorithms.
The simulations confirmed quantitatively the obtained scalings and, most importantly, allowed to characterise the pumping protocol for a realistic system.
In the presence of harmonic confinement, we computed the shift of the centre of mass of the atomic cloud as a diagnostic of pumping.
If experimentally realised, this measurement would constitute a direct observation of a many-body Chern number.

\begin{acknowledgments}
{\it Acknowledgements.} We thank S.~Barbarino, M.~Burrello, M.~Calvanese Strinati, G.~Cappellini,
J.~Catani, M.~Dalmonte, L.~Fallani, L.~Livi, M.~Mancini, G.~Pagano, and C.~Sias for valuable discussions,
and S.~Sinigardi for technical support. We acknowledge INFN-CNAF for providing us
computational resources and support, and D. Cesini in particular.
We acknowledge financial support of EU-IP-SIQS (L.T., D.R., and R.F.), EU-IP-QUIC (R.F),
SNS 2014 Projects (R.F.), QSYNC (R.F.), LabEX ENS-ICFP: ANR-10-LABX-0010/ANR-10-IDEX-0001-02 PSL* (L.M.),
ISF Grant No.~1243/13, and the Marie Curie CIG Grant No.~618188 (ES).
\end{acknowledgments}

\clearpage
\newpage

\clearpage
\setcounter{equation}{0}%
\setcounter{figure}{0}%
\setcounter{table}{0}%
\renewcommand{\thetable}{S\arabic{table}}
\renewcommand{\theequation}{S\arabic{equation}}
\renewcommand{\thefigure}{S\arabic{figure}}

\onecolumngrid

\begin{center}
  {\Large Supplemental Material for: \\ \mytitle}

\vspace*{0.5cm}
Luca Taddia, Eyal Cornfeld, Davide Rossini, Leonardo Mazza, Eran Sela, and Rosario Fazio
\vspace{0.25cm}
\end{center}

\section{APPENDIX A:  RESULTS OF the NEXT-TO-ADIABATIC EXPANSION}

In this appendix we explicitly derive the adiabatically pumped charge and the next-to-adiabatic leading order for the oscillating corrections in Eqs.~\eqref{QR} and~\eqref{QWZ} of the main text.

We follow the results of Refs.~\cite{SOrtiz2008, SOrtiz2014} and expand the time dependent
wavefunction $|\Psi^h(s)\rangle \simeq |\Psi^h(s)\rangle_0 + |\Psi^h(s)\rangle_1 +...,$
with respect to the pumping period $T$ such that $|\Psi^h(s)\rangle_n =  \mathcal{O}(T^{-n})$:
\begin{equation}
\begin{gathered}
	\label{apt01}
	|\Psi^h(s)\rangle_0=e^{-iT\omega_0(s)}\sum_{h'}[U^0(s)]_{hh'}|0^{h'}(s)\rangle;\\
	\begin{split}
	|\Psi^h(s)\rangle_1=
	\frac{i}{T}\sum_{n>0}\sum_{h_n}e^{-iT\omega_0(s)}\frac{[U^0(s)M^{0n}(s)]_{hh_n }}{E_n(s)-E_0(s)}|n^{h_n}(s)\rangle
	-\frac{i}{T}\sum_{n>0}\sum_{h_n}e^{-iT\omega_n(s)}\frac{[M^{0n}(0)U^n(s)]_{hh_n }}{E_n(0)-E_0(0)}|n^{h_n}(s)\rangle
	\\-\frac{i}{T}\sum_{h'}\left(\sum_{n>0}\int_0^s ds'\frac{[U^0(s') M^{0n}(s') M^{n0}(s') U^0(s')^\dagger]_{hh'}}{E_n(s')-E_0(s')}\right)|\Psi^{h'}(s)\rangle_0;
	\end{split}	
\end{gathered}	
\end{equation}
where $[M^{mn}(s)]_{h_m h_n}=\langle n^{h_n}(s)|\partial_s m^{h_m}(s)\rangle$.
The instantaneous wavefunction, which is followed by the system in the adiabatic limit, $|\Psi^h(s)\rangle_0$, as well as the first order $T^{-1}$ correction to the wavefunction,
$|\Psi^h(s)\rangle_1$, depend on the Wilczek-Zee (WZ) matrices:
\begin{equation}
	[U^n]_{hh'}=[\mathcal{T}e^{-\int_0^s M^{nn}(s')ds'}]_{hh'}.
\end{equation}

We first evaluate the adiabatic contribution to the charge which is pumped by $\Phi_S(s)=2\pi s$ in the real direction:
\begin{equation}\label{Qdef}
	Q_R^{(h)}=T\int_0^{1} ds\langle  \Psi^h(s)|\partial_{\Phi_R}\hat{H}(s)|\Psi^h(s)\rangle.
\end{equation}
It originates from the non-oscillatory matrix elements of ${}_0\langle\Psi|\partial_{\Phi_R}\hat{H}|\Psi\rangle_1$
and depends on the WZ matrices:
\begin{equation}
\begin{aligned}
Q_R^{(h)}&=-i\int_0^1 ds\sum_{h'h''}[U^0(s)^\dagger]_{h'h}[U^0(s)]_{hh''}\sum_{n>0}\sum_{h_n}\frac{\langle n^{h_n}(s)|\partial_s \hat{H}|0^{h''}(s)\rangle\langle 0^{h'}(s)|\partial_{\Phi_R} \hat{H}|n^{h_n}(s)\rangle}{(E_{n}(s)-E_{0}(s))^2}+\mathrm{H.c.}\\
&=-i\int_0^{1} ds\sum_{h'h''}\langle \partial_{\Phi_R} 0^{h'}|[U^0(s)^\dagger]_{h'h}[U^0(s)]_{hh''}|\partial_s 0^{h''}\rangle+\mathrm{H.c.}.
\end{aligned}
\end{equation}
By averaging the contributions from $q$ ground states, we obtain Eq.~\eqref{QR}:
\begin{equation}
	Q_R=\frac{1}{q}\sum_h Q_R^{(h)}=i\int_0^{2\pi} d\Phi_S\frac{1}{q}\sum_h[\langle \partial_{\Phi_S} 0^h|\partial_{\Phi_R} 0^h\rangle-\langle \partial_{\Phi_R} 0^h|\partial_{\Phi_S} 0^h\rangle].
\end{equation}

We next evaluate the next-to-adiabatic correction, $\delta Q_R$, which arises from
the oscillatory matrix elements of ${}_0\langle\Psi|\partial_{\Phi_R}\hat{H}|\Psi\rangle_1$.
These matrix elements oscillate as $e^{-i T (\omega_n(s) - \omega_0(s))}$, and hence vanish at $T \to \infty$.
The non-adiabatic correction can be determined using the following theorem:
\begin{equation}
\int_0^1 f(s)e^{iT g(s)}ds=\frac{1}{iT}\left[\frac{f(1)}{\dot{g}(1)}e^{iT g(1)}-\frac{f(0)}{\dot{g}(0)}e^{iT g(0)}\right]+\mathcal{O}(T^{-2}).
\end{equation}
It directly yields the behaviour depicted in Eq.~\eqref{Tvar:text}:
\begin{equation}\label{Tvar}
	\delta Q_R = \sum_{n>0}\frac{A_n+B_n\cos(\bar\Delta_{n0}T+\varphi_n)}{\bar{\Delta}_{n0}T},
\end{equation}
where the structure functions $f$, $g$, and thus the constants $A_n$, $B_n$, $\varphi_n$,
are explicitly derived from Eqs.~\eqref{apt01} and~\eqref{Qdef}.

\section{APPENDIX B: Pumping in the synthetic dimension}

In order to numerically validate the finite-size scalings proposed in the section of the main text \textit{Topological fractional pump, we consider the system with periodic boundary conditions.}
The pump is driven by $\Phi_R(\tau)=2\pi\tau/T,\;\tau\in[0,T]$, while $\Phi_S$ is kept constant in time.
The charge is pumped in the synthetic direction, and is defined by
\begin{equation} 
Q_{S,m}(\tau,\Phi_S)=\int_0^\tau d\tau'\left\langle\psi(\tau',\Phi_S)\right|\hat J^{(S)}_m(\Phi_S)\left|\psi(\tau',\Phi_S)\right\rangle;
\qquad
\hat J^{(S)}_m(\Phi_S)=i\Omega\sum_je^{-i\gamma j+i\frac{\Phi_S(\tau)}{2\mathcal{I}+1}}\hat c_{j,m}^\dagger\hat c_{j,m+1}+\mbox{H.c.}
\end{equation}
The situation is different from the case studied in the section of the main text \textit{Numerical results}, because (i) the system has periodic boundary conditions along the synthetic direction, whereas the system studied in the main text has open boundary conditions in the real direction, and (ii) pumping now occurs along the short dimension.

\begin{figure}[h]
  \centering
  \includegraphics[width=.45\linewidth,trim=0.4cm 0cm 2.3cm 0.6cm, clip=true]{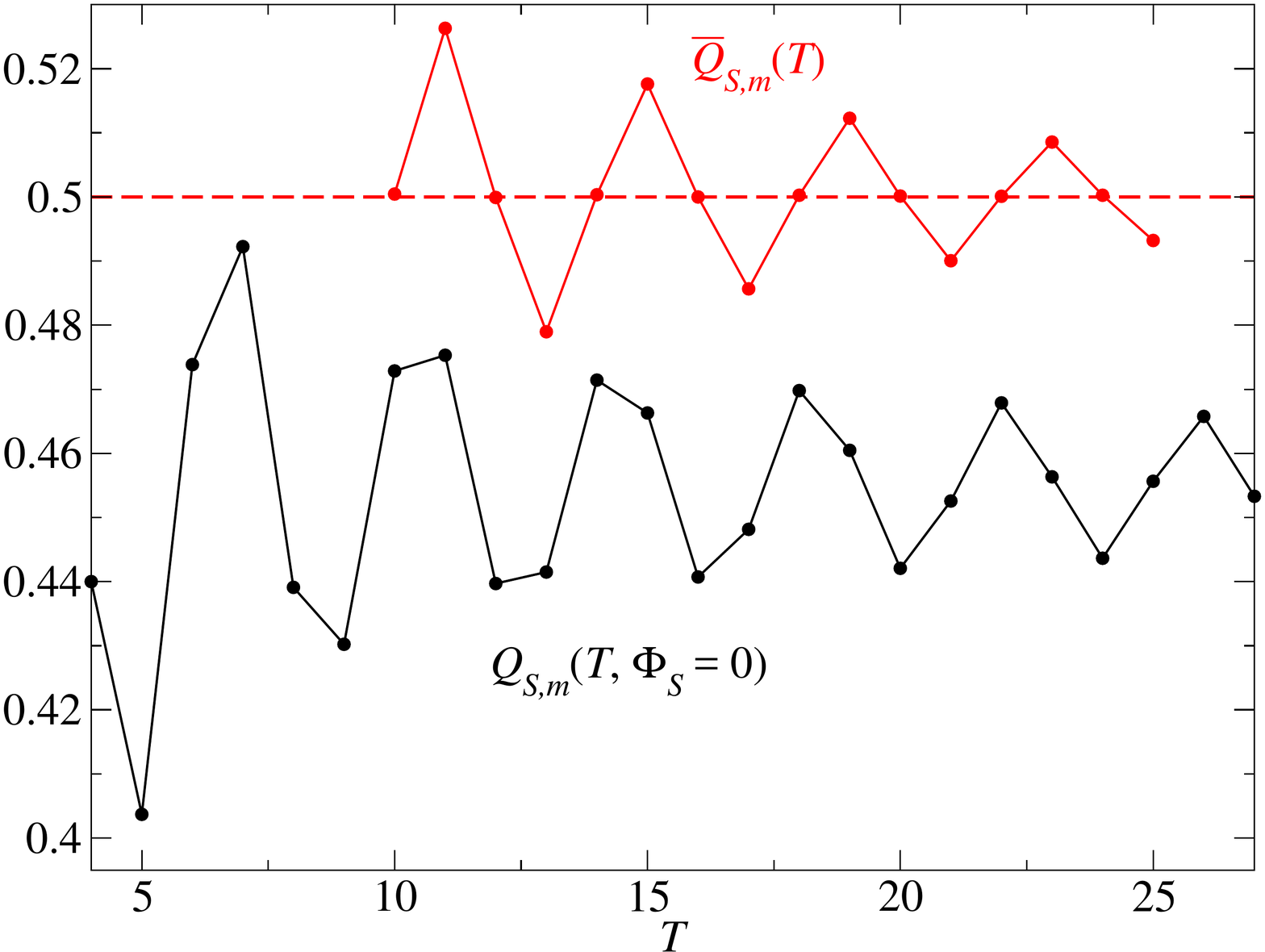}
  \caption{Charge transported in a period by the toroidal pump: the value for $\Phi_S = 0$, $Q_{S,m}(\tau,\Phi_S)$ (black line), and the averaged value over several values of $\Phi_S$, $\bar Q_{S,m} \equiv Q_{S,m,av}(T)$ (red line). The average is obtained
    by sampling the interval $[0,2\pi]$ with 11 values of $\Phi_R$, and is checked
    to be independent on such sampling if dense enough (not shown).
    }
  \label{FStor}
\end{figure}

According to our analytical study, also this protocol supports a quantised pumped charge, but can suffer from an important finite-size effect, due to the intrinsic smallness of the synthetic dimension:
\begin{equation}
 Q_S = \bar Q + \mathcal O ((2 \mathcal I+1)^{-1});
\end{equation}
where $\bar Q_S$ is the pumped charge averaged over $\Phi_S$.
This is shown in Fig.~\ref{FStor} for $\nu=1/2$ and $\Phi_S=0$ (black line): compared to the pump along the real direction, for the same values of the parameters the pumped charge
significantly deviates from the exact fractional value.
Such deviation can be eliminated by considering several constant non-vanishing values of $\Phi_S\in[0,2\pi]$ and averaging over them:
\begin{equation}
\bar Q_{S,m}(T)=\int_0^{2\pi} \frac{d\Phi_S}{2 \pi} Q_{S,m}(T,\Phi_S).
\end{equation}
This is shown in Fig.~\ref{FStor} (red line): the pumped charge, when averaged over a sufficiently large number of values of $\Phi_S$, displays damped oscillations around $1/2$ as a function of $T$.

\section{APPENDIX C: Exact mapping to composite particles.}

\subsection{Single wire}

We present here details on the exactly-solvable model employed in the main text, which was already introduced, although in slightly different forms, in Refs.~\cite{SSela2011, SCornfeld2015}.
For simplicity, we begin by considering a one-dimensional lattice of length $L$ with open boundary conditions and with $N$ spinless fermions; they interact with nearest-neighbour repulsion:
\begin{equation}
 \hat H = -t \sum_{j=1}^{L-1} \left[ \hat c^\dagger_j \hat c_{j+1} + \text{H.c.} \right] + U \sum_{j=1}^{L-1} \hat n_j \hat n_{j+1}.
\end{equation}

In the hard-core limit $U \to \infty$, there is a one-to-one correspondence between the accessible Hilbert-space of this model and that of $N$ effective fermions moving in a lattice of length $L' = L-N+1$.
In order to define the Fock basis for the two spaces, we introduce the fermionic operators $\hat d_j^{(\dagger)}$ which satisfy canonical anticommutation relations and represent the effective fermions. The Fock states which define the basis that we are going to use are labelled by  vectors of length $N$, $\vec \alpha$ and $\vec \beta$, which indicate the position of the $N$ particles (we use the convention that $\alpha_1 < \alpha_2 < \ldots < \alpha_n$; similarly for $\vec \beta$). These states are given by:
\begin{equation}
 | \vec \alpha \rangle = \hat c_{\alpha_N}^\dagger  \ldots c_{\alpha_2}^\dagger  \hat c_{\alpha_1}^\dagger |0\rangle ;
 \qquad
 | \vec \beta \rangle = \hat d_{\beta_N}^\dagger \ldots \hat d_{\beta_2}^\dagger \hat d_{\beta_1}^\dagger |0\rangle .
\end{equation}
 The vector $\vec \alpha$ takes values in $\{1, 2, \ldots L \}$ and satisfies the hard-core constraint, whereas the vector $\vec \beta$ takes values in $\{1, 2, \ldots L'  \}$ and does not satisfy any constraint.
Each particle in the reduced lattice corresponds to one particle and one empty site to its right on the original lattice.
The relation between the $\vec \alpha$ and $\vec \beta$ vectors is given by:
\begin{equation}
 \alpha_1 = \beta_1; \quad
 \alpha_2 = \beta_2+1; \quad
 \alpha_3 = \beta_3+2; \quad
 \ldots  \qquad 
 \alpha_m = \beta_m+m-1.
\label{eq:app:labeling}
\end{equation}

Assuming that both the $\hat c_j^{(\dagger)}$ and the $\hat d_j^{(\dagger)}$ are fermionic operators, we now show that the representation of the kinetic operators:
\begin{equation}
 \hat H_{\rm kin} = \hat H_{\mathrm{kin},l}+\text{H.c.};
 \quad 
 \hat H_{\mathrm{kin},l} =-t \sum_{j=1}^{L-1}  \hat c^\dagger_j \hat c_{j+1} ;
 \qquad 
 \hat H'_{\mathrm{kin}} = \hat H'_{\mathrm{kin},l}+\text{H.c.}; \quad 
 \hat H_{\mathrm{kin},l}'=  -t \sum_{j=1}^{L'-1}  \hat d^\dagger_j \hat d_{j+1} ;
\end{equation}
are identical in the two basis. We focus on $\hat H_{\mathrm{kin},l}$ and $\hat H_{\mathrm{kin},l}'$; the same analysis holds for the Hermitian conjugate.
Let us take two states, $| \vec \alpha \rangle$ and $|\vec \alpha' \rangle$, and compute the matrix element: $\langle  \vec \alpha' | \hat H_{\mathrm{kin},l} |  \vec \alpha \rangle$.
We can envision two possibilities: (i) $\vec \alpha$ and $\vec \alpha'$ differ for only one element, or (ii) $\vec \alpha$ and $\vec \alpha'$ are equal or differ for more than one element.
We stress that the comparison should be done element by element.

We begin by studying the latter case, for the matrix element is zero.
Let us denote $| \vec \beta \rangle$ and $| \vec \beta' \rangle$ the Fock states corresponding to $| \vec \alpha \rangle$ and $| \vec \alpha' \rangle$. In this case, the matrix element $\langle  \vec \beta' | \hat H_{\mathrm{kin},l}' |  \vec \beta \rangle$ is obtained by looking at Eq.~\eqref{eq:app:labeling}: one can see that $| \vec \beta \rangle$ and $| \vec \beta' \rangle$ are either equal or differing by more than one element. Thus, $\langle  \vec \beta' | \hat H_{\mathrm{kin},l}' |  \vec \beta  \rangle = 0$.

The former case is more complicated. Let us assume that $\vec \alpha$ and $\vec \alpha'$ differ only because of one single element, $\alpha_n$ and $\alpha'_n$. First of all, the matrix element is again zero if $\alpha_n'-\alpha_n \neq -1$ (the operator describes the motion of one site to the left). This property is easily reflected in the language of the fictitious fermions.
We now consider what happens when $\alpha_n'-\alpha_n = -1$, i.e. when the matrix element is different from zero. First of all, we notice that the matrix element is different from zero also in the language of the fictitious fermions, because $\beta_n'-\beta_n = -1$.
Importantly, the matrix element is equal because it is always equal to $-t$. The operators $\hat H_{\mathrm{kin},l}$ and $\hat H_{\mathrm{kin},l}'$ are both bosonic and thus commute with the list of $\hat c_j^\dagger$ or $\hat d_j^\dagger$ through which they should be passed to turn the $|  \vec \alpha  \rangle$ ($|  \vec \beta  \rangle$) into the $|  \vec \alpha'  \rangle$ ($|  \vec \beta'  \rangle$) state.

Since the matrix representations of $\hat H_{\mathrm{kin}}$ and $\hat H_{\mathrm{kin}}'$ in the basis are similar, we can freely choose to work with the real fermions or with the fictitious fermions, as long as we are interested in the spectral properties of the model.

\subsection{Ladder}

We consider the generic case of a $n$-leg ladder of length $L$ with open boundary conditions; the system is populated by $N$ interacting spinless fermions.
The model Hamiltonian is $\hat H = \hat H_0 + \hat H_1 + \hat H_{\rm int}$, where: 
\begin{subequations}
\begin{align} 
\hat H_0 =& - t \sum_{j=1}^{L-1}\sum_{m} \left[ \hat c_{j,m}^\dagger \hat c_{j+1,m} + \text{H.c.} \right]; \\
\hat H_{1} =& \Omega \sum_{j,m} \left[e^{i \gamma j} \hat c_{j,m+1}^\dagger \hat c_{j,m} + \text{H.c.} \right],
\quad \mathcal I+1 \equiv - \mathcal I; \\
\hat H_{\rm int} =& \sum_{r \geq 0} V(r) \sum_{j=1}^{L-r} \hat n_j \hat n_{j+r}, \quad \hat n_{j} = \sum_m \hat c^\dagger_{j,m} \hat c_{j,m}.
\end{align}
\end{subequations}
The term $\hat H_0$ describes the hopping within each wire ($m = -\mathcal I, \ldots \mathcal I$), and the inter-chain hopping is described by $\hat H_1$, where $\gamma$ is the magnetic flux per plaquette. In this model, the interaction potential is independent of the rung indexes, but only depends on the linear distance $r$. Similar to the Hubbard model, the solely dependence of the interaction on the total density makes the model $\hat H_0 + \hat H_{\rm int}$ invariant with respect to SU($2 \mathcal I+1$) rotations of the spinor $(\hat c_{j,-\mathcal I} , \ldots,  \hat c_{j, \mathcal I})^{T}$.

We specialise to an interaction potential that vanishes beyond the interaction range $\xi$:
\begin{equation} 
V (r) =  
U  \text{ for } r \leq \xi, 
\qquad
V(r) =  0 \text{ for } r > \xi;
\end{equation}
and we focus on the regime $U \to + \infty$.
In this hard-core limit, the interaction becomes a
constraint: states containing two particles which are horizontally separated by $\xi$ rungs or less acquire a very high energy $\mathcal O(U )$.
Similar to what was discussed in the simplest previous case, the allowed
states for $N$ fermionic particles on the ladder of length $L$ and open boundary conditions are in one to one correspondence with the states of a constrained model. 
This model consists of $N$ fictitious particles on a ladder
of reduced length $L' = L  -(N - 1)\xi$ subject to the additional constraint of not having two fictitious particles on the
same rung. 
Each particle in the reduced lattice corresponds to one particle and $\xi$ empty rungs to its right on the original lattice.

The part $\hat H_0 + \hat H_{\rm int}$ takes the shape of a Hamiltonian with only on-site interaction. We introduce fermionic operators $\hat d_{j,m}$ and write:
\begin{equation}
\hat H_0 + \hat H_{\rm int} \; \to \;
\hat H_0' + \hat H'_{\rm int} =
- t \sum_{j=1}^{L'-1} \sum_m \left[ \hat d_{j,m}^\dagger \hat d_{j+1,m} + \text{H.c.} \right]+
U \sum_{j} \hat \nu_j \hat \nu_{j}; \quad \hat \nu_{j} = \sum_m \hat d^\dagger_{j,m} \hat d_{j,m}.
\end{equation}
This result is the generalisation of the result presented in the previous section, where we focused on a single wire.

Since the site $j$ in the new lattice correspond to the location $j + \xi \sum_{k=1}^{j-1} \hat \nu_k$ in the original lattice, the inter-chain coupling becomes
\begin{equation}
\hat H_{1} \; \to \; \Omega \sum_{j,m} \left[e^{i \gamma \left[ j + \xi \sum_{k=1}^{j-1} \hat \nu_k\right]} \hat d_{j,m+1}^\dagger \hat d_{j,m} + \text{H.c.} \right],
\quad \mathcal I+1 \equiv - \mathcal I; 
\end{equation}
which is non-local. 
However, the non-locality disappears for special values of the flux
\begin{equation}
 \gamma = \frac{2 \pi a}{\xi}; \qquad a \in \mathbb N.
\end{equation}
In this case, the new particles are subject to the same value of the flux, $\gamma' = \gamma$. The original Hamiltonian has been mapped to a new formally-identical Hamiltonian with $L'$ sites and hard-core interactions $\xi ' = 0$. 
For a given filling $\nu$, the density of the original particles, $n$, and that of the new particles, $n'$, are:
\begin{equation}
 n = \nu (2 \mathcal I +1)\frac{\gamma}{2 \pi  } = \nu (2 \mathcal I +1) \frac{ a}{\xi};
 \qquad \qquad 
 n' = \frac{N}{L'} = \frac{1}{n^{-1} - \xi \frac{N-1}{N}}
 \; \xrightarrow{\text{thermodynamic limit}} \;
 \frac{1}{n^{-1}-\xi}.
\end{equation}
The new filling factor is thus:
\begin{equation}
 \nu' = n' \frac{2 \pi}{(2 \mathcal I +1) \gamma} = \frac{1}{\nu^{-1}- (2 \mathcal I +1) a \frac{N-1}{N}} \; \xrightarrow{\text{thermodynamic limit}} \;
 \frac{1}{\nu^{-1}- (2 \mathcal I +1) a }.
\end{equation}

As an example, we can set $\mathcal I = 1/2$, $a = 1$, $\xi = 3$, $N = 50$ and we get:
\begin{align}
 & \nu = \frac 13, \quad \xi = 3, \quad \gamma = \frac{2 \pi}3, \quad L = 225, \quad n = \frac{2}{9}; \nonumber \\
 & \nu' =  0.96 \; \xrightarrow{\text{thermodynamic limit}} \;1, \quad \xi' = 0, \quad \gamma' = \frac{2 \pi}3, \quad L' = 78, \quad n' = 0.64 \; \xrightarrow{\text{thermodynamic limit}} \; \frac 23. \nonumber 
\end{align}
This example demonstrates that the properties of the $\nu' = 1$ system with on-site interactions, $\xi' = 0$, can be related to the properties of the fractional phase $\nu = \frac{1}{3}$ of the original particles.
Notice that the state $\nu = 1/3$, which has an expected quasi-degeneracy of $3$, is not directly mapped to the state $\nu = 1$, which is not degenerate. The slight correction scaling as $1/N$, which makes $\nu' = 0.96$ and not $\nu' = 1$, may be responsible for the accurate description of the quasi-degeneracies occurring in the original system.
This is a problem which requires a thorough analysis with many-body numerical methods and goes beyond the scope of this article and supplemental material.

Before concluding this section, we remark that a similar mapping can be obtained for bosons as well, by including a Jordan-Wigner transformation. 

\section{APPENDIX D: Energy gaps for the $\nu = 1$ state in presence of on-site repulsion}

We present an exact-diagonalisation calculation of the gap of a three-leg ladder system with $\nu = 1$. The on-site interaction term is raised from $U=0$ (the free system can be analytically shown to have a gap) to $U = 10 t$. We observe the presence of a constant and saturating gap which substantiate the claim that raising $U$ to the hard-core limit, the system does not encounter any phase transition and thus many-body topological invariants can be computed for the free system.

\begin{figure}[h]
  \centering
  \includegraphics[width=.45\linewidth]{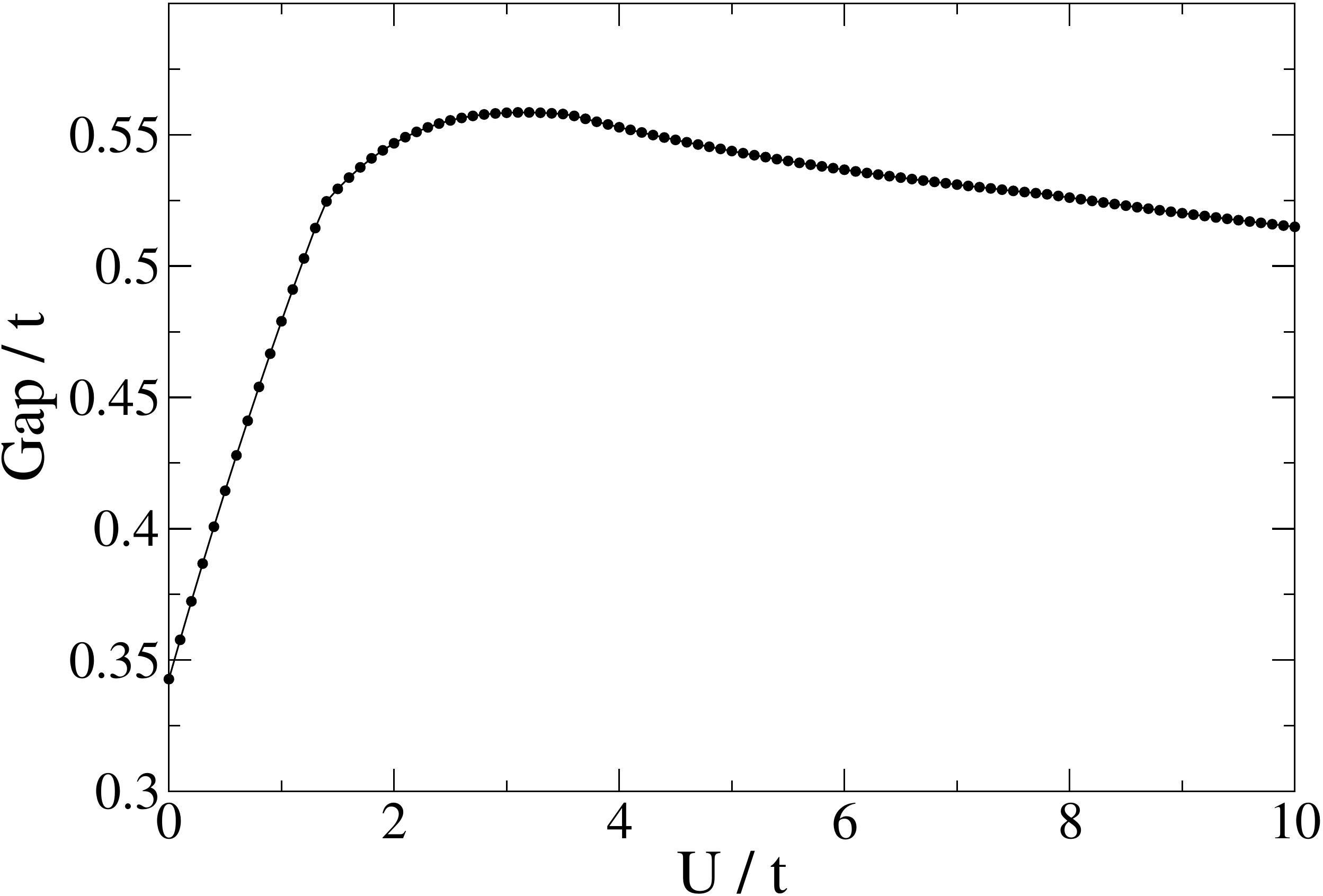}
  \caption{Exact diagonalisation calculation of the energy gap for a system with periodic boundary conditions, $L = 12$, $N = 6$, $\gamma = \pi /3$, $\Omega = 0.3 t$. The system has open boundary conditions in the real direction and periodic boundary conditions in the synthetic direction.
    }
  \label{Gaps}
\end{figure}

\section{APPENDIX E: Evaluation of the Pumped charge using the mapping   from fractional to integer filling in the reduced lattice model}

In this appendix we explicitly derive the relation between the pumped charge in the reduced lattice to the pumped charge of the fractional model.

The reduced lattice described in the main text has length $L'=L/q$ and is thus subjected
to the fluxes $\Phi'_R=\Phi_R \frac{L'}{L}$, $\Phi'_S=\Phi_S$.
We evaluate the contribution of a non-degenerate ground state to Eq.~\eqref{QWZ},
\begin{equation}\label{2piq}
	\bar{Q}=\frac{i}{2\pi}\int_0^{\frac{2\pi}{q}} d\Phi'_R\int_0^{2\pi} d\Phi'_S[\langle \partial_{\Phi'_S} 0|\partial_{\Phi'_R} 0\rangle-\langle \partial_{\Phi'_R} 0|\partial_{\Phi'_S} 0\rangle].
\end{equation}
This is not an integral over a closed surface and is hence not integer-valued.
However, as a consequence of the symmetry $\Phi_R\to\Phi_R+2\pi$, we may perform
an extended pump sequence $\Phi_R(\tau)=q 2\pi \tau/T$.
This sequence must pump $q$ times the charge of the original pump, hence yielding 
\begin{equation}
	\bar{Q}=\frac{i}{q} \oiint \frac{d^2\Phi'}{2\pi}[\langle \partial_{\Phi'_S} 0|\partial_{\Phi'_R} 0\rangle-\langle \partial_{\Phi'_R} 0|\partial_{\Phi'_S} 0\rangle]=\frac{\bar Q'}{q}.
\end{equation}


\begin{thebibliography}{99}
\bibitem{footnote0}
	We disregard for simplicity the effect of Clebsch-Gordan coefficients and take $\Omega$ independent from $m$.
\bibitem{footnote}
	The equality $\bar{Q}= \bar Q' / q$ holds when the density is lesser than
        the particle-hole-symmetry density, i.e., $q  n=n'<1/2$,
        whereby by direct calculation one has $\bar{Q}'=p$.
\end{thebibliography}
\end{document}